# Determination of the Equation of State of Dense Matter

Pawel Danielewicz[1,2], Roy Lacey[3] & William G. Lynch[1*]

**Nuclear collisions can compress nuclear matter to densities achieved within neutron stars and within core-collapse supernovae. These dense states of matter exist momentarily before expanding. We analyzed the flow of matter to extract pressures in excess of $10^{34}$ pascals, the highest recorded under laboratory-controlled conditions. Using these analyses, we rule out strongly repulsive nuclear equations of state from relativistic mean field theory and weakly repulsive equations of state with phase transitions at densities less than three times that of stable nuclei, but not equations of state softened at higher densities because of a transformation to quark matter.**

The nucleon-nucleon interaction is generally attractive at nucleon-nucleon separations of 1 to 2 fm ($1 \times 10^{-13}$ cm to $2 \times 10^{-13}$ cm) but becomes repulsive at small separations ( $< 0.5$ fm) making nuclear matter difficult to compress. As a consequence, most stable nuclei are at approximately the same "saturation" density, $\rho_0 \approx 2.7 \times 10^{14}$ g/cm$^3$, in their interiors, and higher densities do not occur naturally on Earth. Matter at densities of up to $\rho = 9\rho_0$ may be present in the interiors of neutron stars (1), and matter at densities up to about $\rho = 4\rho_0$ may be present in the core collapse of type II supernovae (2). The relationship between pressure, density, and temperature described by the equation of state (EOS) of dense matter governs the compression achieved in supernovae and neutron stars as well as their internal structure and many other basic properties (1-5). Models that extrapolate the EOS from the properties of nuclei near their normal density and from nucleon-nucleon scattering are commonly exploited to study such dense systems (1, 3-9). Consequently, it is important to test these extrapolations with laboratory measurements of high-density matter.

[1] National Superconducting Cyclotron Laboratory and Department of Physics and Astronomy, Michigan State University, East Lansing MI 48824-1321, USA.
[2] Gesellschaft für Schwerionenforschung, 64291 Darmstadt, Germany.
[3] Department of Chemistry, State University of New York, Stony Brook NY 11794-3400, USA.
[*] To whom correspondence should be addressed. E-mail: lynch@nscl.msu.edu

Nuclear collisions provide the only means to compress nuclear matter to high density within a laboratory environment. The pressures that result from the high densities achieved during such collisions strongly influence the motion of ejected matter and provide the sensitivity to the EOS that is needed for its determination (10-19). Full equilibrium is often not achieved in nuclear collisions. Therefore, it is necessary to study experimental observables that are associated with the motions of the ejected matter and to describe them theoretically with a dynamical theory (20-27).

To relate the experimental observables to the EOS and the other microscopic sources of pressure, we apply a model formulated within relativistic Landau theory, which includes both stable and excited (delta, N*) nucleons (that is, baryons) as well as pions (20, 28). It describes the motion of these particles by predicting the time evolution of the (Wigner) one-body phase space distribution functions $f(\mathbf{r}, \mathbf{p}, t)$ for these particles using a set of Boltzmann equations of the form:

$$\frac{\partial f}{\partial t} + (\nabla_{\mathbf{p}} \boldsymbol{e}) \cdot (\nabla_{\mathbf{r}} f) - (\nabla_r \boldsymbol{e}) \cdot (\nabla_p f) = I \,. \tag{1}$$

In this expression, $f(\mathbf{r}, \mathbf{p}, t)$ can be viewed semi-classically as the probability of finding a particle, at time t, with momentum $\mathbf{p}$ at position $\mathbf{r}$. The single particle energies ε in Eq. 1 are given in a local frame by

$$\boldsymbol{e} = KE + U \,, \tag{2}$$

where $KE$ the kinetic energy and $U$ is the average (mean field) potential, which depends on the position and the momentum of the particle and is computed self-consistently using the distribution functions $f(\mathbf{r}, \mathbf{p}, t)$ that satisfy Eq. 1 (20, 28). The particle density is $\mathbf{r}(\mathbf{r}, t) = \int d\mathbf{p} \cdot f(\mathbf{r}, \mathbf{p}, t)$; the energy density $e$ can be similarly computed from $\mathbf{e}$ and $f(\mathbf{r}, \mathbf{p}, t)$ by carefully avoiding an over-counting of potential energy contributions.

The collision integral I on the right hand side of Eq. 1 governs the modifications of $f(\mathbf{r}, \mathbf{p}, t)$ by elastic and inelastic two body collisions caused by short-range residual interactions (20, 28). The motions of particles reflect a complex interplay between such collisions and the density and momentum dependence of the mean fields. Experimental measurements (12-19, 29-31), theoretical innovations, and detailed analyses (10, 20-29, 32-34) have all provided important insights into the sensitivity of various observables to two-body collisions (29, 32) and the density and momentum dependence (28, 33, 34) of the mean fields. The present work builds upon these earlier pioneering efforts.

**Compression and expansion dynamics in energetic nucleus-nucleus collisions.**

Collision dynamics play an important role in studies of the EOS. Several aspects of these dynamics are illustrated in Fig. 1 for a collision between two Au nuclei at an incident kinetic energy of 2 GeV per nucleon (394 GeV). The observables sensitive to the EOS are chiefly related to the flow of particles from the high-density region in directions perpendicular (transverse) to the beam axis. This flow is initially zero but grows with time as the density grows and pressure gradients develop in directions transverse to the beam axis. The pressure can be calculated in the equilibrium limit by taking the partial derivative of the energy density $e$ with respect to the baryon (primarily nucleon) density ρ:

$$P = \mathbf{r}^2 \cdot \left( \frac{\partial (e/\mathbf{r})}{\partial \mathbf{r}} \right)_{s/\mathbf{r}}, \tag{3}$$

at constant entropy per nucleon $s/\mathbf{r}$ in the colliding system. The pressure developed in the simulated collisions (Fig.1) is computed microscopically from the pressure-stress tensor $T^{ij}$, which is the non-equilibrium analogue of the pressure [see supporting on-line material (SOM)]. Different theoretical formulations concerning the energy density would lead to different pressures (that is, to different EOSs for nuclear matter) in the equilibrium limit, in these simulations and in the actual collisions.

At an elapsed time of $3\times10^{-23}$ s in the reaction, the central density (in Fig. 1b') exceeds $3\rho_0$. The corresponding back panel, labeled (b), indicates a central pressure greater than 90 MeV/fm$^3$ (1 MeV/fm$^3$ = $1.6\times10^{32}$ Pa; that is $1.6\times10^{27}$ atmospheres.). These densities and pressures are achieved by inertial confinement; the incoming matter from both projectile and target is mixed and compressed in the high-density region where the two nuclei overlap. Participant nucleons from the projectile and target which follow small impact parameter trajectories (at $x,y \approx 0$), contribute to this mixture by smashing into the compressed region compressing it further. The calculated transverse pressure in the central region reaches ~80% of its equilibrium value after ~$4\times10^{-23}$ sec (Fig. 1c') and is equilibrated for the later times in Fig. 1. Equilibrium is lost at even later times, but only after the flow dynamics are essentially complete.

Spectator nucleons, which are those that avoid the central region by following large impact parameter trajectories (with large $|x|$ >6 fm), initially block the escape of compressed matter along trajectories in the reaction plane and force the matter to flow out of the compressed region in directions perpendicular to the reaction plane (Fig. 1, b to d) Later, after these spectator nucleons pass, nucleons from the compressed central region preferentially escape along in-plane trajectories parallel to the reaction plane that are no longer blocked. This enhancement of in-plane emission is beginning to occur to a limited extent in Fig. 1e at this incident energy of 2 GeV per nucleon. This later in-plane emission becomes the dominant direction at higher incident energies of 5 GeV per nucleon, where the passage time is considerably less. Thus, emission first develops out of plane (along the $y$ axis in Fig. 1) and then spreads into all directions in the $x$-$y$ plane.

The achievement of high densities and pressures, coupled with their impact on the motions of ejected particles, provide the sensitivity of collision measurements to the EOS. The directions in which matter expands and flows away from the compressed region depend primarily on the time scale for the blockage of emission in the reaction plane by the spectator matter and the time scale for the expansion of the compressed matter near $x \approx y \approx z \approx 0$. The blockage time scale can be approximated by $2R/(\gamma_{cm}v_{cm})$, where $R/\gamma_{cm}$ is the Lorentz contracted nuclear radius, and $v_{cm}$ and $\gamma_{cm}$ are the incident nucleon velocity and the Lorentz factor, respectively, in the center-of-mass reference frame. The blockage time scale therefore decreases monotonically with the incident velocity. The expansion time scale can be approximated by $R/c_s$ where $c_s = c\sqrt{\partial P/\partial e}$ is the sound velocity in the compressed matter and $c$ is the velocity of light. The expansion time scale therefore depends (via $c_s$) on the energy density $e$ and upon the nuclear mean field potential $U$ according to Eqs. 2 and 3 and the associated discussion. This provides sensitivity to the density dependence of the mean field potential, which is important because uncertainties in the density dependence of the mean field make a dominant contribution to the uncertainty in the EOS. More repulsive mean fields lead to higher pressures and to a more rapid expansion when the spectator matter is still present. This causes preferential emission perpendicular to the reaction plane where particles can escape unimpeded. Less repulsive mean fields lead to slower expansion and preferential emission in the reaction plane after the spectators have passed.

**Analyses of EOS-dependent observables.**

The comparison of in plane to out-of-plane emission rates provides an EOS dependent experimental observable commonly referred to as elliptic flow. The sideways deflection of spectator nucleons within the reaction plane due to the pressure of the compressed region, provides another observable. This sideways deflection or transverse flow of the spectator fragments occurs primarily while the spectator fragments are adjacent to the compressed region, as shown in Fig.1b' to 1d'. The velocity arrows in Fig. 1d' and 1e' suggest that the changes in the nucleon momenta that result from a sideways deflection are not large. However, these changes can be extracted precisely from the analysis of emitted particles (31). In general, larger deflections are expected for more repulsive mean fields, which generate larger pressures, and conversely, smaller deflections are expected for less repulsive ones.

In terms of the coordinate system in Fig.1, matter to the right (positive $x$) of the compressed zone, originating primarily from the projectile, is deflected along the positive $x$ direction; and the matter to the left, from the target, is deflected to the negative $x$ direction. Experimentally, one distinguishes spectator matter from the projectile and the target by measuring its rapidity $y$, a quantity that in the nonrelativistic limit reduces to the velocity component $v_z$ along the beam axis (35). For increasing values of the rapidity, the mean value of the $x$ component of the transverse momentum increases monotonically (12, 14-16, 31). Denoting this mean transverse momentum as $<p_x>$ and corresponding transverse momentum per nucleon in the detected particle as $<p_x/A>$, we find that larger values for the pressure in the compressed zone due to more repulsive EOSs lead to larger values for the directed transverse flow F defined (12) by

$$F = \frac{d \left\langle p_x / A \right\rangle}{d \left( y / y_{cm} \right)} \bigg|_{y / y_{cm} = 1} ,\qquad (4)$$

where $y_{cm}$ is the rapidity of particles at rest in the center of mass and $A$ is the number of nucleons in the detected particle. (F can be viewed qualitatively as the tangent of the mean angle of deflection in the reaction plane. Larger values for F correspond to larger deflections.) The open and solid points in Fig. 2 show measured values for the directed transverse flow in collisions of $^{197}$Au projectile and target nuclei at incident kinetic energies $E_{beam}/A$ ranging from about 0.15 to 10 GeV per nucleon (29.6 to 1970 GeV total beam kinetic energies) and at impact parameters of $b$=5 to 7 fm ($5 \times 10^{-13}$ cm to $7 \times 10^{-13}$ cm) (13-16). The scale at the top of this figure provides theoretical estimates for the maximum densities achieved at selected incident energies. The maximum density increases with incident energy; the flow data are most strongly influenced by pressures corresponding to densities that are somewhat less than these maximum values.

The data in Fig. 2 display a broad maximum centered at an incident energy of about 2 GeV per nucleon. The short dashed curve labeled "cascade" show results for the transverse flow predicted by Eq. 1, in which the mean field is neglected. The disagreement of this curve with the data shows that a repulsive mean field at high density is needed to reproduce these experimental results. The other curves correspond to predictions using Eq. 1 and mean field potentials of the form

$$U = \left( a\boldsymbol{r} + b\boldsymbol{r}^{\boldsymbol{n}} \right) / \left[ 1 + \left( 0.4\,\boldsymbol{r} \, / \, \boldsymbol{r}_0 \right)^{\boldsymbol{n}-1} \right] + \boldsymbol{d}U_p \,. \qquad (5)$$

Here, the constants $a$, $b$ and $\boldsymbol{n}$ are chosen to reproduce the binding energy and the saturation density of normal nuclear matter while providing different dependencies on density at much higher density values, and $\boldsymbol{d}U_p$ describes the momentum dependence of the mean field potential (28, 33, 34) (see SOM text). These curves are labeled by the curvature $K \equiv 9dp/d\boldsymbol{r}|_{s/r}$ of each EOS about the saturation density $\rho_0$. Calculations with larger values of K, for the mean fields above, generate larger transverse flows, because those mean fields generate higher pressures at high density. The precise values for the pressure at high density depend on the exact form chosen for U. To illustrate the dependence of pressure on K for these EOSs, we show the pressure for zero temperature symmetric matter predicted by the EOSs with K=210 and 300 MeV in Fig. 3. The EOS with K=300 MeV generates about 60% more pressure than the one with K=210 MeV at densities of 2 to 5$\rho_0$ (Fig.3).

Complementary information can be obtained from the elliptic flow or azimuthal anisotropy (in-plane versus out-of-plane emission) for protons (24, 25, 36). This is quantified by measuring the average value <cos2φ>, where φ is the azimuthal angle of the proton momentum relative to the x-axis defined in Fig. 1. (Here, tanφ= $p_y/p_x$ , where $p_x$ and $p_y$ are the in-plane and out-of-plane components of the momentum perpendicular to the beam.) Experimental determinations of <cos2φ> include particles that, in the center-of-mass frame, have small values for the rapidity y and move mainly in directions perpendicular to the beam axis. Negative values for <cos2φ> indicate that more protons are emitted out-of-plane (φ≈90°or φ≈270°) than in-plane (φ≈0°or φ≈180°), and positive values for <cos2φ> indicate the reverse situation.

Experimental values for <cos2φ> for incident kinetic energies $E_{beam}$/A ranging from 0.4 to 10 GeV per nucleon (78.8 to 1970 GeV total beam kinetic energies) and impact parameters of b = 5 to 7 fm (5-7×$10^{-13}$ cm) (17-19) are shown in Fig. 4. Negative values for <cos2φ>, reflecting a preferential out-of-plane emission, are observed at energies below 4 GeV/A indicating that the compressed region expands while the spectator matter is present and blocks the in-plane emission. Positive values for <cos2φ>, reflecting a preferential in-plane emission, are observed at higher incident energies, indicating that the expansion occurs after the spectator matter has passed the compressed zone. The curves in Fig. 4 indicate predictions for several different EOSs. Calculations without a mean field, labeled "cascade," provide the most positive values for <cos2φ>. More repulsive, higher pressure EOSs with larger values of K provide more negative values for <cos2φ> at incident energies below 5 GeV per nucleon, reflecting a faster expansion and more blocking by the spectator matter while it is present.

Transverse and elliptic flows are also influenced by the momentum dependencies $dU_p$ of the nuclear mean fields and the scattering by the residual interaction within the collision term I indicated in Eq. 1. Experimental observables such as the values for <cos2φ> measured for peripheral collisions, where matter is compressed only weakly and is far from equilibrated (28), now provide significant constraints on the momentum dependence of the mean fields (21, 28). This is discussed further in the SOM (see SOM text). The available data (30) constrain the mean-field momentum dependence up to a density of about $2\rho_0$. For the calculated results shown in Figs. 2 to 4, we use the momentum dependence characterized by an effective mass $m*=0.7m_N$, where $m_N$ is the free nucleon mass, and we extrapolate this dependence to still higher densities. We also make density dependent in-medium modifications to the free nucleon cross-sections following Danielewicz (28, 32) and constrain these modifications using observables sensitive to stopping in collisions, such as the longitudinal momentum distributions ($p_z$ distributions) of reaction products.

**Sensitivity to the pressure and to the symmetric matter EOS.**

The elliptic and transverse flow observables are sensitive to the mean field and to the EOS at central densities of 2 to $5\rho_0$ (Figs. 2 and 4). We compared the two observables to the calculations, and did not find a unique formulation of the EOS that reproduces all of the data. At incident energies of 2 to 6 GeV/A, for example, the transverse flow data lie near or somewhat below (to the low pressure side of) the $K$=210 MeV calculations, whereas the elliptic flow data lie closer to the $K$=300 MeV calculations. Some discrepancies are also observed between the two sets of experimental transverse data at incident energies of 0.25 to 0.8 GeV/A. Although it is not possible to fully resolve the inconsistencies between theory and experiment at the present time, one can still use these data to provide constraints on the EOS. For example, calculations without a mean field (cascade) or with a weakly repulsive mean field ($K$=167 MeV) provide too little pressure to reproduce either flow observable at higher incident energies (and correspondingly higher densities). The calculations with $K$=167 MeV and $K$=380 MeV provide lower and upper bounds on the pressure in the density range $2 \leq \rho/\rho_0 \leq 5$. These comparisons also suggest that the upper bound must lie lower than the pressures corresponding to the $K$=380 MeV curve at most densities and that a field that is less repulsive than $K$=380 MeV at densities $\rho>3\rho_0$ could provide a stricter upper bound on the pressure.

Our transport theory calculations provide a calibration for the transverse and elliptic flow "barometers." These can be used, in turn, to assess the pressures achieved in the hot and non-equilibrium environment of a nuclear collision. From our transport theory, we determined that maximum pressures in the range of P = 80 to 130 MeV/fm$^3$ ($1.3\times10^{34}$ to $2.1\times10^{34}$ Pa) and P = 210 to 350 MeV/fm$^3$ ($3.4\times10^{34}$ to $5.6\times10^{34}$ Pa) are achieved at incident energies of 2 and 6 GeV per nucleon. These pressures are approximately 23 orders of magnitude larger than the maximum pressures recorded previously under laboratory-controlled conditions (37). They are about 19 orders of magnitude larger than pressures within the core of the sun, but are comparable to pressures within neutron stars.

**Determination of constraints and comparison to theoretical EOSs.**

Comparing the calculations and data of Figs. 2 and 4 and factoring in the uncertainties due to the momentum dependencies of the mean fields and the collision integral, we have assessed the range of pressure-density relationships. These are shown for zero temperature matter and densities of $2 < \rho/\rho_0 < 4.6$ by the shaded region in Fig. 3. These bounds on the EOS for symmetric nuclear matter are the main achievement of this work.

To illustrate the value of these constraints, a few representative theoretical EOS's are shown in Fig. 3. The EOS of Akmal *et al.* (3), which passes through the allowed region, represents a class of models that take the two-nucleon interactions from fits to nucleon-nucleon scattering data. The EOS of Lalazissis *et al.* (RMF:NL3) (6) represents a class of relativistic mean field theory models that derive the nucleon-nucleon interaction from the exchange of effective $\omega$ and $\sigma$ mesons. Although most such models provide too much pressure, we note that a recent inclusion by Typel and Wolter (7) of nonlinear terms in the Lagrangian can reduce the pressure in such models so as to be consistent with the present experimental constraints. The EOS of Boguta (38) illustrates the softening of the EOS that might occur if there were another phase more stable than nuclear matter for densities of about $3\rho_0$. This EOS and the calculation without a mean field produce too little pressure. On the other hand, an EOS that first increases in pressure, consistent with our constraints, such as the Akmal EOS, and then remains constant with density above $\rho/\rho_0 = 3$, consistent with the existence of a different, more stable, phase at higher densities $\rho/\rho_0 > 4$, such as transition to a phase composed of quarks and gluons (the quark-gluon plasma), cannot be precluded by the present analysis.

Our constraints on the EOS of symmetric matter are relevant to the dynamics of supernovae and to the properties of neutron stars, where such densities are achieved (1). Supernovae involve admixtures of neutrons and protons that are similar to the Au + Au system; the application of these constraints to supernovae is more straightforward than is the application of these constraints to extremely neutron-rich environments such as neutron matter or neutron stars. In such neutron rich environments, one must consider how the EOS depends on the difference between the neutron and proton concentrations. This concentration difference vanishes for symmetric matter, but in pure neutron matter gives rise to an additional source of pressure $P_{sym} = \rho^2 d(E_{sym}(\rho))/d\rho \mid_{s/\rho}$, which depends on the symmetry energy $E_{sym}(\rho)$. The symmetry energy determines how the energies of nuclei and nuclear matter depend on the difference between neutron and proton densities. This energy is repulsive and is the reason why light nuclei have nearly equal numbers of protons and neutrons. Few experimental constraints on the density dependence of $E_{sym}(\rho)$ exist. Therefore, we employ, in the following, the two parameterizations for $E_{sym}(\rho)$ with the weakest (Asy_soft) and strongest (Asy_stiff) density dependence proposed by Prakash *et al.* (4) to assess the sensitivity of neutron star properties to the asymmetry term. This assessment is summarized in Fig. 5.

Assuming either an asymmetry term (Asy_stiff) with strong density dependence or an asymmetry term (Asy_soft) with weak density dependence (see SOM text), the allowed regions from Fig. 3 can be extrapolated (Fig. 5). Clearly, the uncertainty in the pressure due to the asymmetry term, represented by the difference between the pressures for these two "allowed" regions in Fig. 5, exceeds the remaining uncertainty in the pressure due to the symmetric matter EOS, represented by the width of each region. The pressure in the actual neutron star environment is somewhat smaller than that for neutron matter (Fig. 5), reflecting the small fraction of nucleons that are protons. The precise values of this proton fraction and many other static and dynamical properties of these dense astrophysical objects depend on the density dependence of the asymmetry term (4).

In comparison to these "allowed" pressures, the pressure due to the Fermi motion of a pure neutron gas (Fermi gas) is comparatively small; the remaining pressure must arise from the repulsive mean field potential. The EOS of Akmal et al. (3) and the av14uv11 EOS of Wiringa (8) are both models that take the two-nucleon interactions from fits to nucleon-nucleon scattering data. The EOS [MS($\zeta$=0,$\xi$=0)] of Müller and Serot *et al.* (9) represents a class of relativistic mean field theory models that derive the nucleon-nucleon interaction from the exchange of effective $\omega$ and $\sigma$ mesons. Its prediction is essentially the same as the neutron matter predictions for the model of Lalazissis *et al.* (6) (RMF:NL3 in Fig. 3). Although these models appear to provide too much pressure, other relativistic mean field theory models of Müller and Serot *et al.* (9) and the calculations of Glendenning *et al.* (5), (GWM:neutrons) predict lower pressures. The uncertainty in the pressure due to the asymmetry term widens the range of possible EOSs that may be consistent with the experimental data. For this purpose, it is important to obtain experimental constraints on the asymmetry term (39-41) (see SOM text) and to complement them with improved constraints on the EOS of symmetric nuclear matter.

We have analyzed the flow of matter in nuclear collisions to determine the pressures attained at densities ranging from two to five times the saturation density of nuclear matter. We obtained constraints on the EOS of symmetric nuclear matter that rule out very repulsive EOSs from relativistic mean field theory and very soft EOSs with a strong phase transition at densities $\rho < 3\rho_0$, but not a softening of the EOS due to a transformation to quark matter at higher densities. Investigations of the asymmetry term of the EOS are important to complement our constraints on the symmetric nuclear matter EOS. Both measurements relevant to the asymmetry term and improved constraints on the EOS for symmetric matter appear feasible; they can provide the experimental basis for constraining the properties of dense neutron-rich matter and dense astrophysical objects such as neutron stars.

**Supporting Online Material**

www.sciencemag.org/cgi/content/full/1078070/DC1

SOM Text

References

Fig. S1

FIGURE CAPTIONS:

Fig. 1. Overview of the dynamics for a Au + Au collision. Time increases from left to right, the center of mass is at $\mathbf{r} = \mathbf{0}$, and the orientation of the axes is the same throughout the figure. The trajectories of projectile and target nuclei are displaced relative to a "head-on" collision by an impact parameter of $b = 6$ fm ($6\times10^{-13}$ cm). The three-dimensional surfaces (middle panel) correspond to contours of a constant density $\rho$ ~ 0.1 $\rho_0$. The magenta arrows indicate the initial velocities of the projectile and target (left panel) and the velocities of projectile and target remnants following trajectories that avoid the collision (other panels). The bottom panels show contours of constant density in the reaction plane (the $x$-$z$ plane). The outer edge corresponds to a density of 0.1 $\rho_0$, and the color changes indicate steps in density of 0.5 $\rho_0$. The back panels show contours of constant transverse pressure in the $x$-$y$ plane. The outer edge indicates the edge of the matter distribution, where the pressure is essentially zero, and the color changes indicate steps in pressure of 15 MeV/fm$^3$ (1 MeV/fm$^3$ = 1.6$\times10^{32}$ Pa; that is ~1.6$\times10^{27}$ atmospheres). The black arrows in both the bottom and the back panels indicate the average velocities of nucleons at selected points in the $x$-$z$ plane and $x$-$y$ planes, respectively.

Fig. 2. Transverse flow results. The solid and open points show experimental values for the transverse flow as a function of the incident energy per nucleon. The labels "Plastic Ball," "EOS," "E877" and "E895" denote data taken from Gustafsson *et al*. (13), Partlan *et al*. (14), Barrette *et al*. (15), and Liu *et al*. (16), respectively. The various lines are the transport theory predictions for the transverse flow discussed in the text. $\mathbf{r}_{max}$ is the typical maximum density achieved in simulations at the respective energy.

Fig. 3. Zero-temperature EOS for symmetric nuclear matter. The shaded region corresponds to the region of pressures consistent with the experimental flow data. The various curves and lines show predictions for different symmetric matter EOSs discussed in the text.

Fig. 4. Elliptical flow results. The solid and open points show experimental values for the elliptical flow as a function of the incident energy per nucleon. The labels "Plastic Ball," "EOS," "E895" and "E877" denote the data of Gutbrod *et al.* (17), Pinkenburg *et al.* (18), Pinkenburg *et al.* (18) and Braun-Munzinger and Stachel (19), respectively. The various lines are the transport theory predictions for the elliptical flow discussed in the text.

Fig. 5. Zero temperature EOS for neutron matter. The upper and lower shaded regions correspond to the pressure regions for neutron matter consistent with the experimental flow data after inclusion of the pressures from asymmetry terms with strong and weak density dependences, respectively. The various curves and lines show predictions for different neutron matter EOSs discussed in the text.

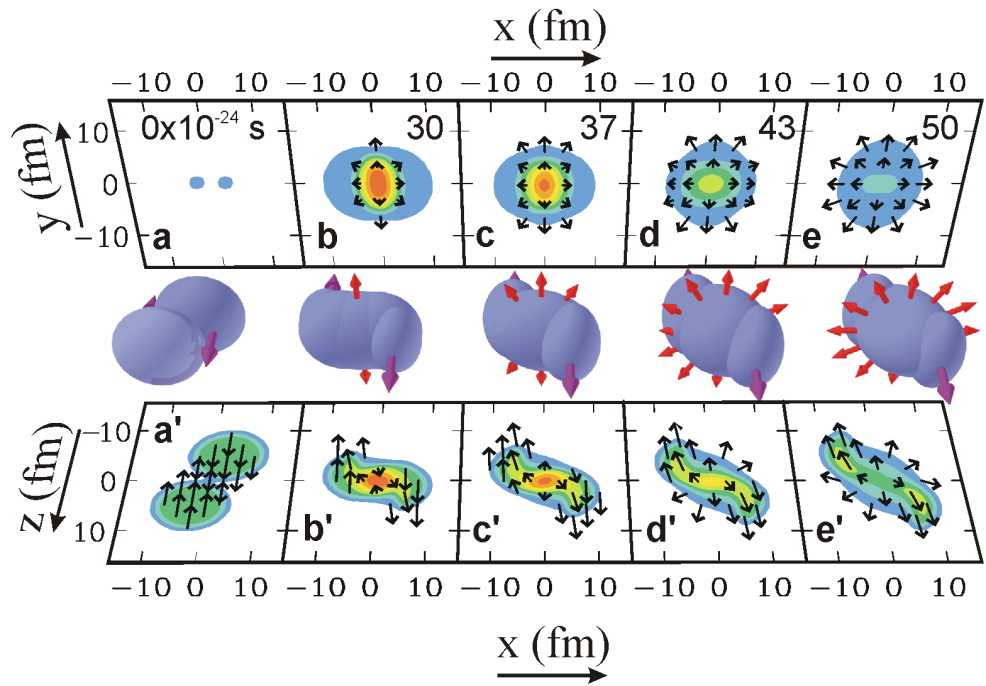

FIGURE 1

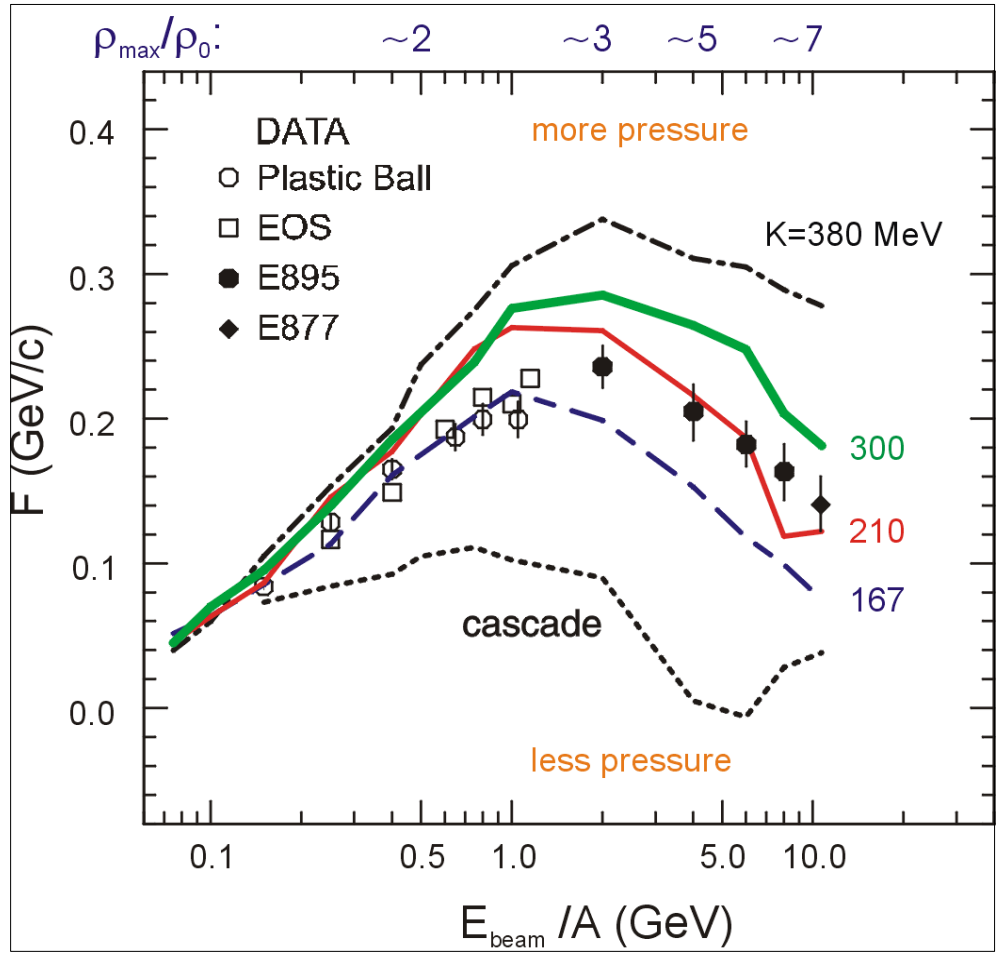

FIGURE 2

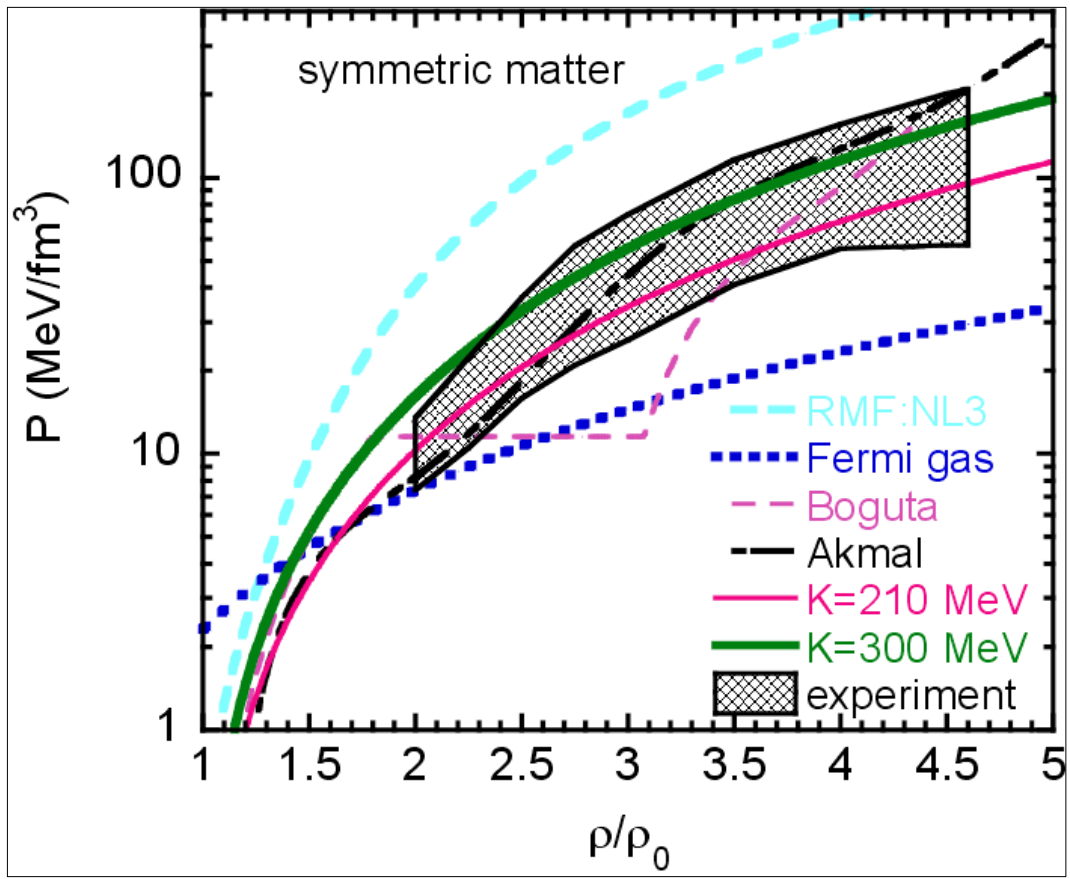

FIGURE 3

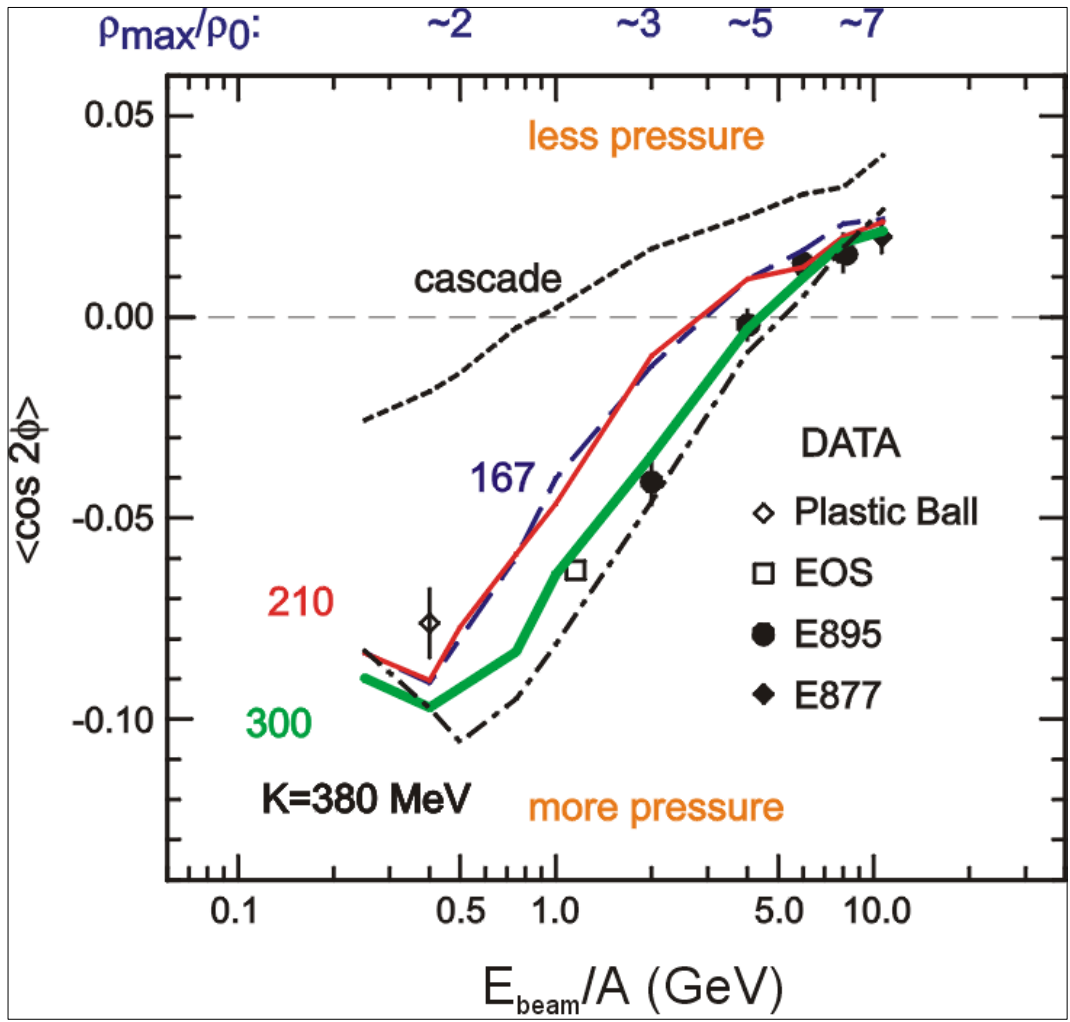

FIGURE 4

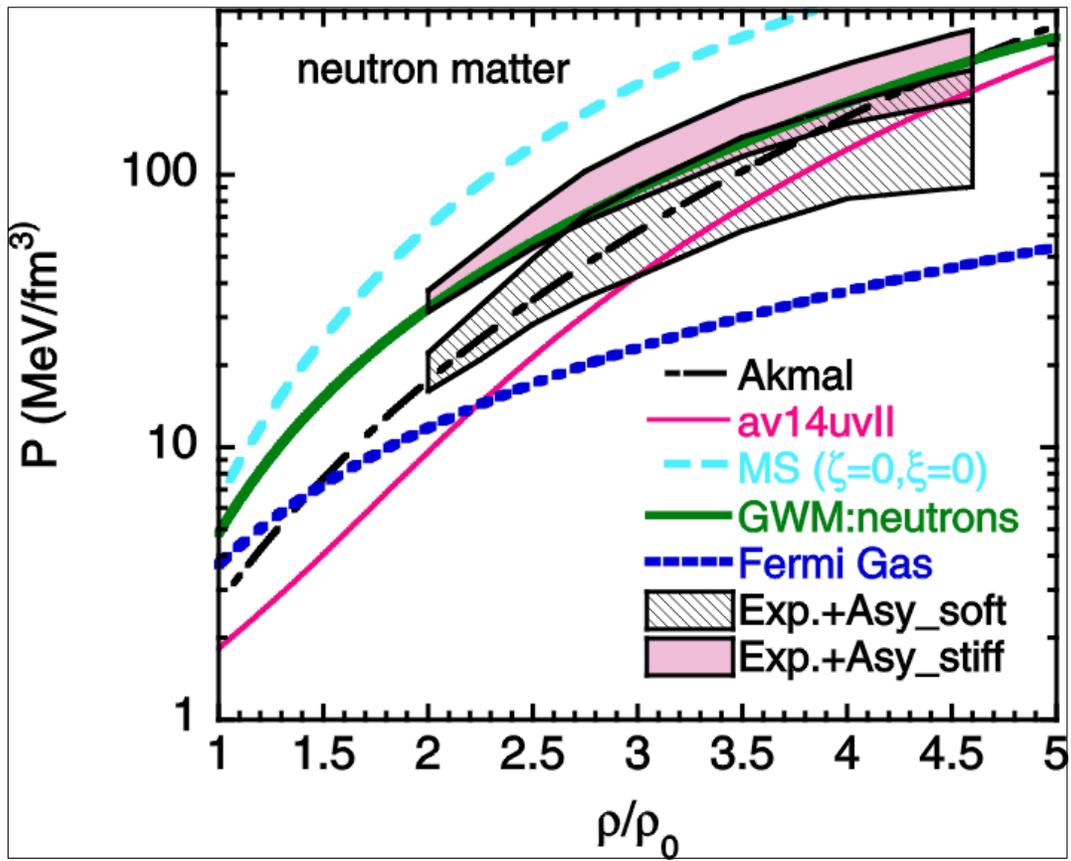

FIGURE 5

# Determination of the Equation of State of Dense Matter

Pawel Danielewicz, Roy Lacey & William G. Lynch

## Supplementary Material

**Pressure calculations:**

The back panels of Fig. 1 of the main article show the mean transverse pressure $\overline{P}_{tr} = (T^{xx} + T^{yy})/2$, where the $T^{ij}$ are components of pressure-stress tensor. The pressure-stress tensor is trivially related to the pressure when equilibrium is assumed, i.e. $T^{ij} = P\delta^{ij}$, where $\delta^{ij}$ is the identity matrix. The collective flow observables analyzed in this paper are primarily sensitive to the pressure driven accelerations in the transverse (x and y) directions, for which the mean transverse pressure is primarily responsible. This connection to the transverse pressure can be deduced from the non-equilibrium Euler equation wherein the acceleration $a^i$ of matter in direction i is given (in its rest frame) by:

$$a^i = -c^2 \partial_j T^{ij} / [e + P'], \qquad (S1)$$

where $P' = a^k a^l T^{kl} / (a^m a^m)$ and repeated indices imply summation over the three Cartesian spatial coordinates.

Eq. S1 contains non-diagonal components $T^{ij}$ of the pressure-stress tensor. However, these turn out to be small at high density compared to the diagonal terms included in $\overline{P}_{tr}$. Notably, the calculated transverse pressure in the central region reaches ~80% of its equilibrium value after $\sim 3 \times 10^{-23}$ sec. It becomes essentially equilibrated after $\sim 4 \times 10^{-23}$ sec (panel **c'**) (i.e. the non-diagonal term are negligible) and remains so for the later times represented in Fig. 1. Equilibrium is lost at later times but only after the flow dynamics are essentially complete. Thus, the collective flow is essentially determined by $\overline{P}_{tr}$.

**Momentum dependence of the mean fields:**

One factor limiting the precision of past analyses of flow data is that both transverse and elliptic flows are also sensitive to the momentum dependencies of the nuclear mean fields and to the uncertainties in the scattering by the residual interaction within the collision term I in Eq 1. Indeed, the global flow observables in Figs. 2 and 4 do not suffice to constrain the density dependence of the mean fields, the momentum dependencies of the nuclear mean fields and the uncertainties in the scattering by the residual interaction without additional data analyses. Thus progress in determining the EOS for dense matter was significantly delayed by ambiguities in interpretation stemming from the sensitivity to the momentum dependence of the mean fields.

The constraints upon the momentum dependence of the mean fields at supranormal densities have recently been obtained through investigations of the momentum dependence of experimental observables (1-3) such as <cos2φ> for peripheral collisions, where matter is less compressed than central collisions. These constraints allow us to assess the momentum and density dependence of the EOS somewhat independently. Supplemental Fig. S1 shows experimental values of <cos2φ> for protons (2) detected at values of the rapidity $0.35 \leq y/y_{beam} \pounds 0.65$, i.e. close to that of the center of mass, as a function of the component, $p_\perp$, of the proton momentum perpendicular to the beam. The various curves correspond to solutions of Eq. 1 (in the main article) for different assumptions about the momentum dependence (1) and they are labeled by the nucleon effective mass corresponding to the assumed momentum dependence. Smaller effective masses lead to a more rapid nucleonic motion, a more rapid emission from the overlap region while the spectator matter is present and, consequently, to more negative values for <cos2φ>. The data (2) constrain the mean-field momentum dependence up to the density of about $2\rho_0$. In the calculations in Figs. 2-4 of the main article, we use the momentum dependence characterized by $m^*=0.7m_N$, where $m_N$ is the free nucleon mass, and extrapolate this dependence to still higher densities. This choice is consistent with nucleon-nucleus scattering. We also make density dependent in-medium modifications to the free nucleon cross-sections following Danielewicz (4) and constrain these modifications using observables characterizing stopping in collisions, including the longitudinal momentum distributions ($p_z$ distributions) of reaction products.

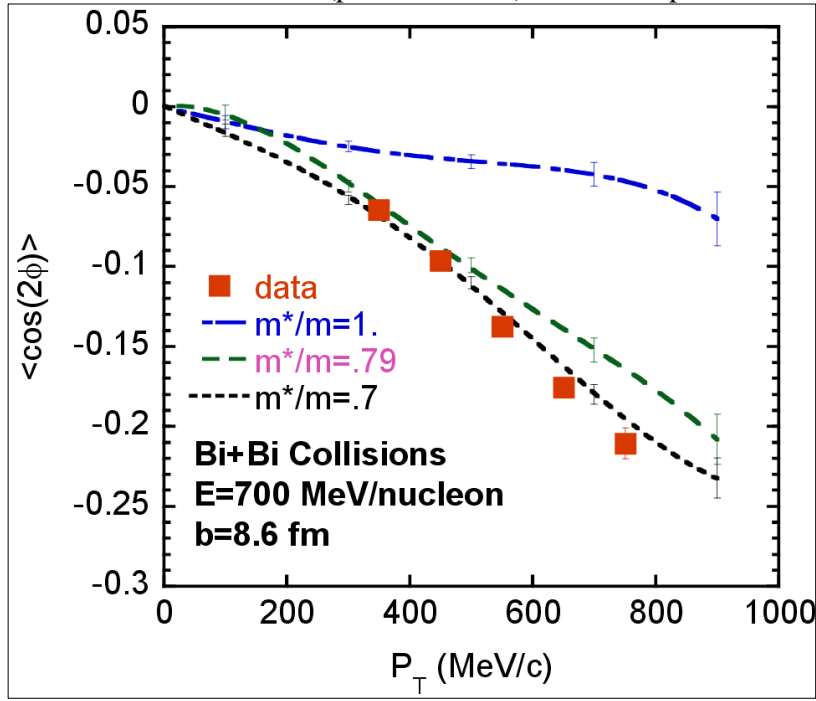

Supplemental Figure 1: Constraints on the momentum dependence of the mean field. The solid points show experimental values (2) for <cos2φ> in Bi+Bi peripheral collisions at incident energies of E =700 MeV/nucleon as a function of the transverse momentum of the detected protons. The various lines are transport theory predictions for <cos2φ> assuming different momentum dependences for the nuclear mean field (1). The error bars on the theoretical curves indicate the statistical uncertainty in the theoretical calculations.

**Proposals for constraining the asymmetry term experimentally:**
Presently, the density dependence of the asymmetry term is poorly constrained by experimental data. Calculations have been performed to identify experimental observables that can provide significant constraints on the density dependence of the asymmetry term (5-9). Li (5), for example, has proposed that the density dependence of the asymmetry term at $\rho > 2\rho_0$ may be probed by comparisons of $\pi^+$ and $\pi^-$ yields or of neutron or proton transverse flows for central collisions between heavy very neutron-rich beams and neutron-rich targets. Experimental tests of these predictions have not yet been performed, but such tests would be well suited to the proposed Rare Isotope Facility (RIA) (6). Calculations also indicate that comparisons of relative proton and neutron emission rates or transverse collective flow values may be sensitive to the density dependence of the asymmetry term at somewhat lower densities (7-10). Measurements of quantities related to the relative proton and neutron emission rates at densities of $\rho < \rho_0$ have actually been reported in refs. (11,12). Brown (13) has also predicted that measurements of the neutron matter radius in $^{208}$Pb could quantitatively probe the asymmetry term at $\rho < \rho_0$.